\documentclass[aps,floats,twocolumn,epsf,prl,showpacs]{revtex4} 

\usepackage{latexsym,amsmath,amssymb,bm,graphicx}
\usepackage{psfrag}
\makeatletter\def\Dated@name{}\makeatother

\newcommand{\deps}{\Delta\varepsilon}

\newcommand{\xna}{x_\text{Na}}

\begin{document}

\title{Enhancement of the Na$_x$CoO$_2$ thermopower due to electronic correlations} 
\author{P. Wissgott$^{1}$, A. Toschi$^{1}$, H. Usui$^{2}$, K. Kuroki$^{2}$ and K. Held$^1$} 
\affiliation{$^1$ Institute for Solid State Physics, Vienna University of Technology, 
1040 Vienna, Austria\\
$^2$ University of Electro-Communications 1-5-1 Chofugaoka, Chofu-shi
Tokyo 182-8585, Japan}
\date{\today}


\begin{abstract} 
 Using the merger of local density approximation and dynamical mean field
theory, we show how electronic
correlations  increase the thermopower of Na$_{0.7}$CoO$_2$ by 200\%. 
The newly revealed mechanism is an asymmetric
shift of (quasi) electrons and holes
away from the Fermi level, concurrent with an asymmetry 
of the respective (group) velocities. 
Exploiting this effect in bandstructure and correlation engineering may lead to a substantial increase of the thermoelectric figure of merit.
\end{abstract} 
 
\maketitle

Climate change and the prospective oil peak necessitate the discovery of new green energy sources. One possibility in this context is to convert hithero unused excess heat into electrical energy using thermoelectrics~\cite{Mahan}. For a real breakthrough and a widespread application of this technology
however, thermoelectric figures of merit $ZT\!\gtrsim\! 3$
are needed~\cite{MRS,solar}. Recently, some  advance was
brought about through phonon \cite{BiTePh} and bandstructure
engineering \cite{PbTeDOS} so that 
$ZT\!\gtrsim\! 1$ could be achieved.
Pushing $ZT$ much higher requires however a thermoelectric 
leap, which  is most likely brought about
by discovering a new class of materials --
at least in the case of~(high-T) superconductivity scientific progress 
went along this line with the discovery
of novel correlated materials.
At commercially available $ZT\!\lesssim \!1$, 
 thermoelectrics are already applied in niche markets
such as radioisotope power systems for
satellites \cite{satellite} where reliability is more important than 
efficiency; and the car industry is taking encouraging
first steps to use thermoelectrical generators \cite{cars}.
The materials  presently used in industry are mainly
Te-based and have been investigated most thoroughly \cite{Handbook}
with respect to optimizing their thermoelectric properties, including
 the aforementioned
 phonon \cite{BiTePh} and bandstructure
engineering \cite{PbTeDOS}. 

 \begin{figure}[t] 
\begin{picture}(300,170)\put(0,0){
\put(30,-5){
\includegraphics[height=7cm]{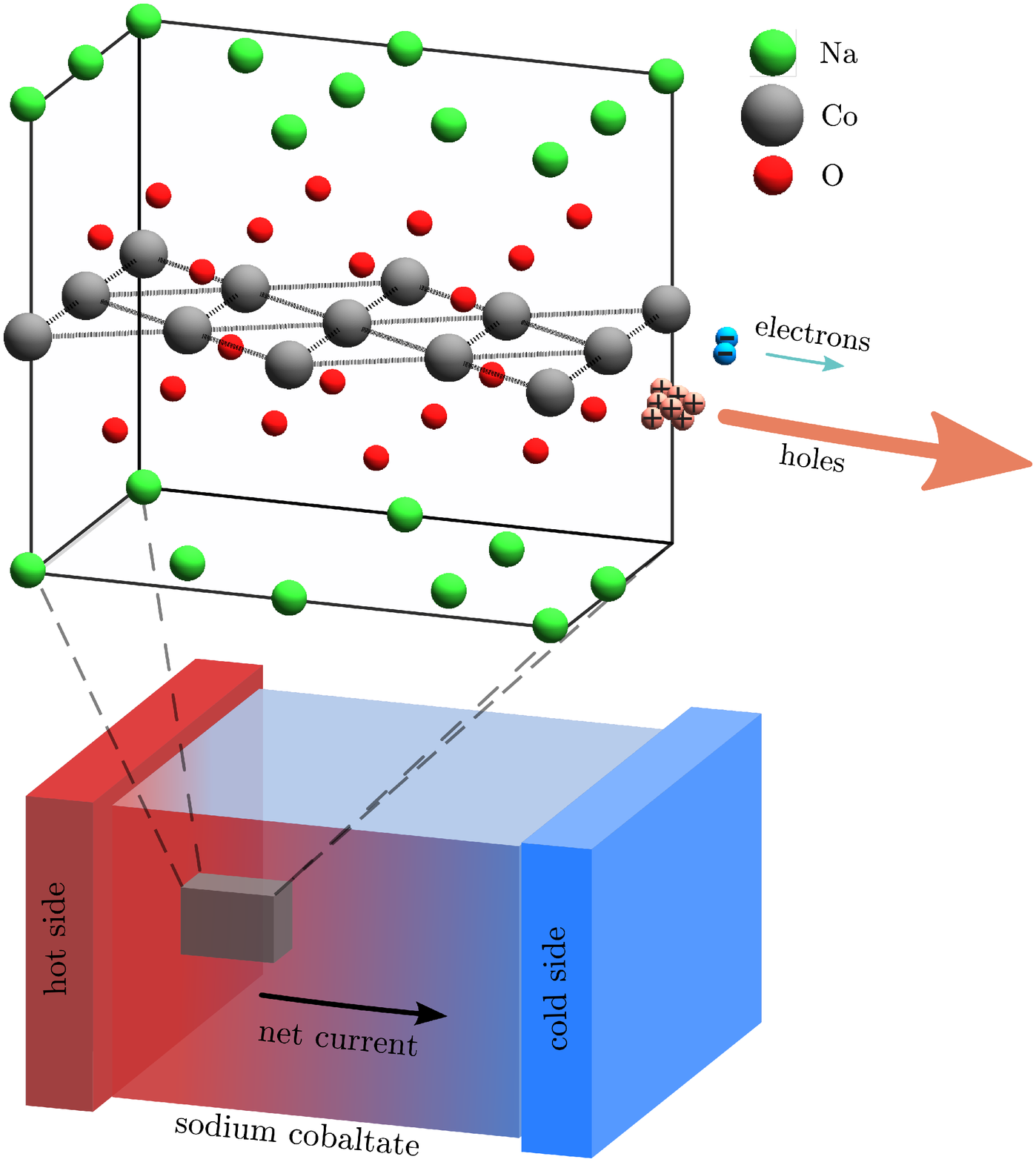}}
}
\end{picture}
\caption{(Color online)~Scheme of  a Na$_{x}$CoO$_2$ thermoelectric module
and its crystal structure. The charge carriers are confined
to the hexagonal layers of Co atoms, which are inequivalent
because of the (disorderly distributed) Na ions 
above and below. \newline
If we apply a heat gradient, electrons and holes alike are
diffusing from the hot to the cold side resulting typically in a very small net current. For a good thermoelectric, we need
however a large net current, i.e. a huge asymmetry between 
electron and hole diffusion. \newline
Due to the pudding-mold bandstructure in Na$_{x}$CoO$_2$, the group
velocity of the holes is already considerably larger than that of the 
electrons, as indicated by the
larger orange~(dark gray) arrow. Electronic correlations additionally shift 
electrons away from the Fermi level, as visualized by the fewer blue electrons 
 and the narrower width of the blue~(light gray) arrow. 
As both effects add up, the electron is much smaller
than the hole current leading to the net current displayed and hence, to the good thermoelectric properties of Na$_{x}$CoO$_2$.
 \label{Fig0} }
\end{figure}

The good thermoelectric properties of Te-based materials are related to their physics of slightly doped semiconductors with a low effective mass and high mobility of the charge carriers\cite{MRS,Handbook}.
A possible improvement are compounds with a pudding-mold type of bandstructure
as those of 
Na$_x$CoO$_2$ \cite{Kuroki} and LiRh$_2$O$_4$\cite{Arita08} -- two oxides with 
remarkably good thermoelectric properties \cite{Terasaki,Lee06,Kaurav09,Okamoto}.

Another route to good thermoelectric properties is through 
electronic correlations \cite{Paschen,transportnanostructures}
which can renormalize bands most substantially, so that narrow quasiparticle bands with a high density of states emerge.
The renormalization factor is most dramatic in  heavy Fermion
systems which have particularly narrow Kondo resonances,  modeled e.g. by a periodic Anderson model
\cite{TEPAM1}. While one might expect the narrowest 
resonances to be most suitable for thermoelectrics, there is a trade-off
since increasing the temperature $T$ above the Kondo temperature 
(essentially the resonance width)  will eventually smear out the resonance itself. 

In this paper, we demonstrate by means of Na$_x$CoO$_2$, which already
has a favorable pudding-mold bandstructure, that the complementary
 effect of electronic correlations can dramatically increase the thermopower,
see Fig.~\ref{Fig0}.
Needed are inequivalent orbitals or sites, so that
electronic correlations can change the ``particle-hole balance'' of the system drastically, shifting more electrons away from the Fermi level than holes.

For treating electronic correlations and bandstructure effects on an equal level, we employ the local density approximation+dynamical mean field theory (LDA+DMFT) approach
\cite{LDADMFT0,LDADMFT1,LDADMFT4} and calculate  thermoelectric response functions as described in \cite{Held09}. Previous LDA+DMFT calculations for Na$_x$CoO$_2$~\cite{Ishida05,Kotliar07,Liebsch08} focused on the spectral function, whereas we naturally aim at  the  transport properties of this material.
Starting point is a tight-binding fit to the LDA bandstructure \cite{Singh} for the Co $a_{1g}$ band 
within the hexagonal CoO$_2$ layers, see Fig.\ \ref{Fig0}. Considering transport properties~(e.g. the thermopower $S$), the
other $d$-bands can be neglected at a
first approximation. In fact,  LDA predicts 
already that these bands are  almost filled, and this result is also 
confirmed experimentally, since $e_g$ hole pockets are absent in angular
resolved photoemission spectroscopy (ARPES)~\cite{Hasan04,Yang04}. 
There are  $x\!<\!1$ Na$^+$ ions per formula unit, which we consider to
be randomly distributed with a Na atom or vacancy in trigonal prismatic coordination to
the Co sites. As it has been
suggested in Ref.\ \onlinecite{Kotliar07}, this gives rise to a  disorder
potential on  the Co sites -- since the positively charged Na$^+$
attracts the Co-electrons.  This disorder potential is of essential
importance for the strong electronic correlations of the
Co $d$-electrons; and as in Ref.\  \onlinecite{Kotliar07},
we take an disorder potential of $\deps\!=\!0.55\,$eV. 
 Our resistivity calculation~(inset of Fig.~\ref{Fig4})
{\em a posteriori} confirms that a disorder potential 
of this strength is present; otherwise the experimental value of the resistivity at room temperature, 
mainly due to the disorder (and not to the electron-electron scattering) 
 could not be explained~\cite{footnote1}. 
Our model Hamiltonian hence reads
\begin{equation}\nonumber
 H\!=\!-\sum_{(ij),\sigma}t_{ij}\ c^{\dagger}_{j\sigma} c_{i\sigma} + U\sum_{ i} c^\dagger_{i\uparrow} c_{i\uparrow}c^\dagger_{i\downarrow}c_{i\downarrow} + \deps\sum_{{i\in \text{vac}},{\sigma}} c^\dagger_{i\sigma} c_{i\sigma},
\end{equation}
 where $c^+_{i\sigma}, c_{i\sigma}$ are creation and annihilation operators of the $a_{1g}$ electrons, with $i$ the site index and $\sigma$ the spin index and t$_{ij}$ denotes the hopping parameter from site $i$ to site $j$, obtained from a  tight-binding fit to LDA data~\cite{Singh,note1}. For the Coulomb repulsion within the $a_{1g}$ orbitals we take a value $U\!=\!3.5\,$eV and for
the Na disorder potential 
  $\deps\!=\!0.55\,$eV as in  \cite{Kotliar07}.
 As a DMFT impurity solver we use
quantum Monte Carlo simulations, where the Na-disorder is included at the level of the coherent potential approximation CPA.
For the calculation of the response functions we proceed
as in~\cite{Held09}, using a Taylor-fit to extract
the low energy behavior of the self energy on the real axis.
\begin{figure}[t] 
\includegraphics[width=6.5cm]{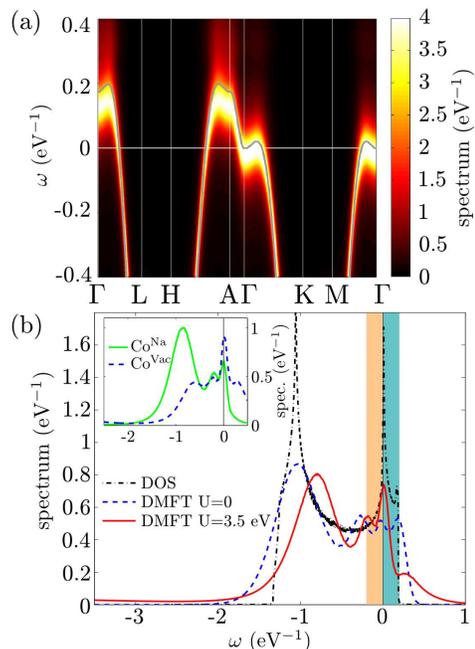} 
\vspace{-0.4cm}
\caption{(Color online) (a) LDA+DMFT k-resolved spectrum of Na$_{0.7}$CoO$_2$ at 290 K
around the chemical potential level~($\omega\equiv 0$), along 
 with the tight-binding band extracted from the LDA calculation (light gray). (b)~LDA+DMFT k-integrated spectra for $T=290$ K  and x=0.7
with and without interaction $U$ and disorder potential $\deps$
(dotted line: U=0, $\deps$=0; dashed line: U=0, $\deps$=0.55 eV; solid line: U=3.5 eV,$\deps$=0.55 eV). For fixed U=0, an increasing disorder potential $\deps$=0$\rightarrow$0.55 eV leads to more symmetric but lower spectral weight in the highlighted interval $\pm 8 k_BT$
around the Fermi level. For fixed $\deps$=0.55 eV, increasing  U=0$\rightarrow$3.5 eV recovers spectral weight on the hole side $\omega\!<\!0$ whereas electrons $\omega\!>\!0$ lose spectral weight in the relevant (highlighted) area. The inset shows the  k-integrated spectra at the two inequivalent lattice sites, with and without an adjacent Na$^+$ (T=290 K, U=3.5 eV, and $\deps$=0.55 eV).} 
\label{Fig:kresolvedspectrum} \label{Fig1} 
\end{figure} 

Fig.\ \ref{Fig1} shows our $k$-resolved  and 
$k$-integrated spectral function.
For example along path $\overline{\rm HA\Gamma K}$ through the Brillouin zone, 
the LDA bandstructure $\varepsilon(k)$ (gray line in \ref{Fig1}(a))
vaguely resembles the form of a pudding mold.
The advantage of the pudding-mold bandstructure is the highly asymmetric group velocity of the electrons and
holes above and below
the Fermi level, respectively. This velocity is given by the slope 
$v_g(k)=\partial \varepsilon(k)/\partial k$  which,
because of the pudding mold,  dramatically changes at the Fermi level at least
in the  $\overline{\rm \Gamma K}$ direction (not so much along $\overline{\rm HA}$).
The LDA density of states, on the other hand, is higher above the Fermi level (for electrons)
than below (for holes), see (Fig.\ \ref{Fig1}(b)). Hence, the asymmetry in $v_g$ which increases the thermopower $S$  is partially compensated by the
 asymmetry  in the density of states. 
In Fig.\ \ref{Fig0} there would be, on the LDA level, more electrons than holes
contributing.\\
The thermoelectric figure of merit is defined as $ZT=S^2/(\rho \kappa) T$, with the temperature $T$, the electrical resistivity $\rho$ and the thermal conductivity $\kappa=\kappa_{el}+\kappa_{ph}$ where $\kappa_{el}$, $\kappa_{ph}$ denote the electronic and phononic contributions, respectively. By use of Kubos formulas in DMFT~\cite{Oudovenko}, one finds $\rho=1/K_0$, $S=-\frac{k_{\rm B}}{T}\frac{{K_1}}{{K_0}}$ and $\kappa_{el}=(K_2-K_1^2/K_0)/T$ with the correlation functions
\begin{align}
K_n\!=\!(-1)^{n+1}\frac{2\pi e^{2-n}}{V}\sum_{k}v_g^2(k)\! \int\! d\omega A^2(k,\omega)  \frac{\partial f}{\partial\omega} \omega^n,\label{Eq:Kn}
\end{align} 
where $V$ is the unit cell volume, $f(\omega)$ is the Fermi-Dirac distribution function. 
Comparing Eq.~(\ref{Eq:Kn}) for $n=0$ and $n=1$, it becomes clear that due to the factor $\omega$ in $K_1$, electron ($\omega>0$) and
hole contributions ($\omega<0$) push $S$ in opposite directions. A high
electron-hole asymmetry is hence mandatory for getting a large $|S|$.

{\small
\begin{table*}[tb]
\begin{tabular}{l c c c c c c}
\hline \hline
parameters & $\rho$(m$\Omega$cm) & $S$($\mu$V/K) & $\kappa_e$(mW/cm K) & $S^2/\rho$($\mu$W/K$^2$ cm) & $Z$ ($10^{-3}$ K) & $ZT$ \\
\hline 
$U=0$, $\deps=0$ &0.21 &  54 & 36.8 & 13.8 & 0.135 & 0.039\\
$U=3.5$ eV, $\deps=0$ &0.31 &  66 & 27.4 & 13.7 & 0.146 & 0.042\\
$U=0$, $\deps=0.55$ eV &2.51 &  25 &  2.9 &  0.2 & 0.004 & 0.001\\
$U=3.5$ eV, $\deps=0.55$ eV &1.37 &  63 &  6.5 &  2.9 & 0.040 & 0.012\\
experiment &1.04\cite{Wang04} &  64\cite{Kaurav09} &  4.0\cite{Sales04} &  3.9\cite{Wang04,Kaurav09} & 0.053\cite{Wang04,Kaurav09,Sales04} & 0.015\cite{Wang04,Kaurav09,Sales04}  \\
\hline \hline
\end{tabular}
\caption{Thermoelectric and transport properties for single crystalline Na$_{0.7}$CoO$_2$ at 290 K.
For the computed figure of merit $Z\!=\!S^2/\rho(\kappa_{ph}+\kappa_{e})$, we take the experimental phononic thermal conductivity
 $\kappa_{ph}\!=66$mW/cm K~\cite{Sales04}. The electronic part of the thermal conductivity $\kappa_{e}$
 is calculated with the correlation functions from Eq.~(\ref{Eq:Kn}) and estimated experimentally by the Wiedemann-Franz law. Note that there are no free parameters and that only disorder~($\deps$) and correlation~($U$) together describe all experimental values reasonably well.}
\label{table1}
\end{table*}
}

Before turning to the electronic correlations, let us first discuss the
effect of disorder  (dashed line of Fig.\ \ref{Fig1}(b)), not yet switching on the interaction $U$. 
As one can see, the spectral weight around the Fermi energy is
smeared out because of the disorder and the spectral weight
is most dramatically reduced. 
As one might expect from 
Eq.\ (\ref{Eq:Kn}), this is very disfavorable for the thermopower. Actually, it is reduced from the disorder-free value \cite{Kuroki} to only S=25 $\mu$V/K without interaction $U$. In Fig.\ \ref{Fig0}, we visualize
this by a narrower arrow for both electron and hole contribution and
a smaller net current.

This radically changes if electronic correlations are taken into account.
Their effect can be best understood by hands of  the inset of Fig.\ \ref{Fig1}(b),
which shows the $k$-integrated Co spectrum locally resolved for
sites with and without an adjacent Na$^+$ ion. The Coulomb interaction
 drive the vacancy-adjacent Co sites more towards half-filling.
Consequently, electronic correlations become stronger 
which reflects in the sharp quasiparticle resonance
and Hubbardesque side bands. In contrast, the
Na$^+$-adjacent Co sites become almost completely filled and are
only weakly correlated. Together, the shift of the Na$^+$-adjacent 
Co spectrum below the Fermi energy and the strong renormalization of the
vacancy-adjacent Co sites yields a higher spectral weight on the hole
than on the electron side. This asymmetry now \emph{adds} to the $v_g(k)$ asymmetry, in stark contrast to the original LDA spectral weight. Note, that the velocities itself are, to first order,  not renormalized by the electronic correlations since these are a property of the underlying LDA orbitals. 
A complete overview of the effect of disorder and correlation on the transport properties is given in Tab.~\ref{table1}.

\begin{figure}[t]
 \centering \includegraphics[width=8 cm]{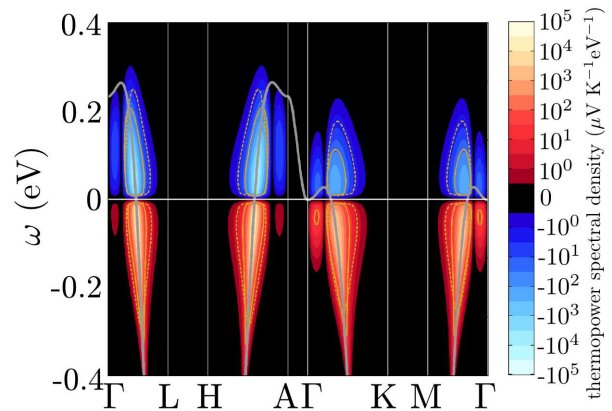}  
\vspace{-0.4cm}
\caption{(Color online) Thermopower spectral density s(k,$\omega$) for $T\!=\!290\,$K, $U\!=\!3.5\,$ eV and $\deps\!=\!0.55\,$eV together with the tight-binding band extracted from the LDA data (gray). Positive values $s>0$~(red for $\omega<0$) contributed by holes dominate negative values $s\!<\!0$~(blue for $\omega>0$) contributed by electrons. The electronic contribution is more sensitive to correlation which is emphasized by the values $s\!=\!\pm 10^2 \!\mu$VK$^{-1}$eV$^{-1}$ highlighted by an isocurve,
comparing $U\!=\!3.5\,$eV (solid) and $U\!=\!0$ (dashed).} 
\label{Fig:thermopowerdensity} \label{Fig3} 
\end{figure} 

To further illustrate the changes induced by correlation,
we introduce a thermopower spectral density
$ s(k,\omega)=-\frac{k_{\rm B}}{T K_0}\frac{2 \pi e}{V} v_g^2(k)\, A^2(k,\omega)\, \frac{\partial f(\omega)}{\partial\omega}\omega
$, 
which integrated over $k$ and $\omega$ yields the thermopower $S$~(Fig. \ref{Fig3}). The contour lines show that, upon (artificially) switching on the proper $U$, thermopower spectral weight is taken away from the electron side. As one can see in Fig. \ref{Fig1}(b) it is 
transferred further away from the Fermi level. In contrast, the reduction of spectral weight on the hole side is much smaller and close to the
$\Gamma$ point $s(k,\omega)$ is even enhanced.  This correlation-induced electron-hole imbalance strongly enhances the thermopower~($S$ increases by $200$\% w.r.t. the case of U=0). 

Finally, we present in Fig.\ \ref{Fig4} the calculated thermopower $S$ as a function of $T$ for various doping; the corresponding
resistivity is shown in the inset. Our LDA+DMFT results agree well with experiment for both quantities including the doping and $T$ dependence.
The large thermopower of Na$_x$CoO$_2$ originates from the pudding mold band structure
and is further enhanced by electronic correlations.
The physical mechanism for the large resistivity on the other hand,
is the disorder scattering.

\begin{figure}[t]
 \includegraphics[width=9cm]{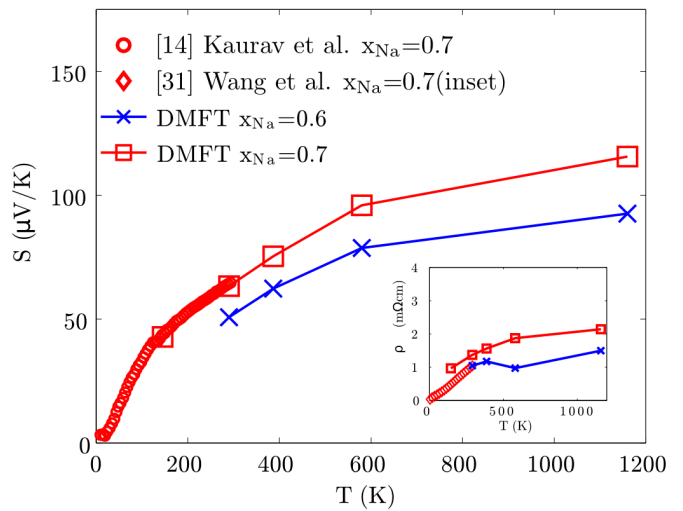}
\caption{(Color online) Thermopower $S$ as a function of temperature for different doping $\xna=0.6,0.7$  showing good agreement to the experiment of Kaurav {\em et al.}~\cite{Kaurav09}. The inset shows the corresponding result for the resistivity compared to experiments by Wang \emph{et al.}~\cite{Wang04}.}\label{Fig:thermopower}\label{Fig4}
\end{figure}

From our calculations we conclude that Na$_x$CoO$_2$ is far from optimal for thermoelectric properties. 
On the bandstructure side, there is still no strong 
electron-hole asymmetry for the group velocities
 in the $\overline{\rm \Gamma{\rm L}}$ and  $\overline{\rm HA}$ direction due to the three dimensionality, see  Fig.\ \ref{Fig1}. Hence, the  electron contribution
(blue in Fig.~\ref{Fig1}(b)) still reduces the thermopower considerably. This can be improved by
bandstructure engineering the pudding mold form, e.g., by
 expanding the c-axis length via intercalation.
Second,  disorder broadens the LDA substantially and dramatically decreases the Seebeck coefficient as well
as the conductivity $\sim1/\rho$. There is hence a high potential for further 
increasing $S$,
e.g., by realizing a similar physical situation on the basis of inequivalent orbitals or  superstructures without disorder. 
Last, but not least, we have shown the crucial role correlation can play, not only because of the narrowing of the coherent excitations but also in strengthening the particle-hole imbalance.
Given the high sensitivity of correlated systems to small changes of the external conditions and recent advances in engineering man-made
heterostructures of correlated materials,
there is plenty of room for improving this correlation effect further.

We can thus conclude from
our study of transport properties in Na$_x$CoO$_2$ 
 that, if properly combined, bandstructure and correlation effects~(as those present in transition metal d and rare earth f orbitals) 
can  considerably improve the  thermoelectric properties 
of materials. 
 Jointly, engineering bandstructure and correlations
hence opens a new route to further enhance S~(and possibly $ZT$) up to the levels required for a widespread technological 
applications.

Financial support from the FWF through GK W004 
and the EU-Indian network MONAMI is gratefully acknowledged.


\onecolumngrid


\begin{thebibliography}{10}
\bibitem{note1}
We obtain $t=(180,-39,-27,0.4)$ meV as nearest- to fourth-nearest neighbor hopping in plane;
$t_{z}=(-18,-5,-2,1,0.5)$ meV are the 
 interplanar parameters from the first (prismatic) Co neighbor up to the fourth.
\end{thebibliography}
\end{document}